\documentclass{article}
\usepackage {graphicx}

\begin{document}

\begin{center}
{\large \bf A Lithium Experiment as the Stringent Test of the
Theory of Stellar Evolution.}

\vskip 0.3in

Anatoly Kopylov and Valery Petukhov\\ Institute of Nuclear
Research of Russian Academy of Sciences \\ 117312 Moscow, Prospect
of 60th Anniversary of October Revolution 7A
\end{center}

\begin{abstract}
We show that a lithium experiment has a potential to confirm or
reject the value 1.5\% of the solar luminosity attributed to a
CNO-cycle by the standard solar model and to prove that the
difference between total energy release of the Sun and what is
produced in a hydrogen sequence is really produced in CNO cycle.
This will be the stringent test of the theory of stellar evolution
and the termination of the long-standing goal -- the neutrino
spectroscopy of the interior of the Sun. At the present time one
can see no other way to solve this task, it can be accomplished
only with a lithium detector utilizing its high sensitivity to
CNO-neutrinos and very high accuracy in the theoretical evaluation
of the cross-section of neutrino capture by $^{7}$Li. The analyses
shows that although a lithium detector is a radiochemical one,
principally it is possible to find separately the fluxes of
$^{13}$N- and $^{15}$O-neutrinos.
\end{abstract}

The primary goal of the solar neutrino experiment since the early
phase of the research pioneered by Raymond Davis was first - to
prove the thermonuclear nature of the energy generation in the
interior of the Sun and second - to find the experimental evidence
that the Sun shines by the pp-chain and not by the CNO cycle. Soon
the goal was formulated \cite{1} as a ``Neutrino Spectroscopy of
the Interior of the Sun'' (NSIS). Since that time the copious
experimental material was accumulated \cite{2}. The field turned
out to be very productive, the experiments not only provided the
first direct evidence of the thermonuclear nature of the energy of
the Sun, they discovered basically new properties of neutrinos --
neutrino oscillations and measured its parameters: $\Delta
$m$^{2}$ and mixing angle \cite{3}. The progress in this field is
really very impressive. In a few years when a lithium project will
be ready for the realization as a full-scale experiment, a lot
more will be accomplished: basically all neutrino fluxes of
hydrogen-sequence reactions and also the neutrino oscillation
parameters will be measured with good accuracy. The only thing
left will be presumably the neutrinos of CNO-cycle (at least now
one can see no way how they could be detected apart from a lithium
experiment). Figure 1 shows CNO reactions in schematic form
\cite{6}.

At the present time only an upper limit of 7.3\% was set \cite{4}
to the fraction of energy that the Sun produces via the CNO fusion
cycle while according to the Standard Solar Model this fraction
constitutes 1.5\% \cite{1}. This limit was set just as a
difference between the total energy release in the Sun found from
the luminosity and the one generated in a hydrogen sequence alone,
found from the fluxes of solar neutrinos measured in solar
neutrino experiments \cite{2} hence it is more like the limit of
the non-hydrogen sequence source of energy. Here it is taken into
account the effect from neutrino oscillations with the oscillation
parameters of the MSW LMA region found in solar neutrino
experiments and KamLAND \cite{5},\cite{5a}.
\begin{figure}[!ht]
\centering
\includegraphics[width=3in]{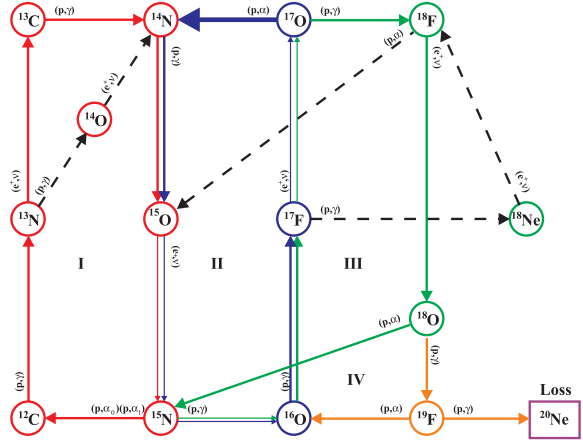}
\caption{The scheme of CNO reactions.}
\end{figure}

\begin{center}
\begin{table}[!b]
Table 1. Standard Model Predictions (BP2000): solar neutrino
fluxes and neutrino capture rates, with 1$\sigma$ uncertainties
from all sources (combined quadratically).\\
\begin{tabular}
{|c|c|c|c|c|} \hline Source& Flux \par
(10$^{10}$cm$^{-2}$s$^{-1}$)& Cl
\par (SNU)& Ga \par (SNU)& Li \par (SNU) \\ \hline pp&
5.95(1.00$^{+0.01}_{-0.01}$)& 0.0& 69.7& 0.0
\\ \hline pep& 1.40$\times $10$^{-2}$(1.00$^{+0.015}_{-0.015}$)& 0.22& 2.8& 9.2
\\ \hline hep& 9.3$\times $10$^{-7}$& 0.04& 0.1& 0.1
\\ \hline $^{7}$Be& 4.77$\times $10$^{-1}$(1.00$^{+0.10}_{-0.10}$)& 1.15& 34.2& 9.1
\\ \hline $^{8}$B& 5.05$\times $10$^{-4}$(1.00$^{+0.20}_{-0.16}$)& 5.76& 12.1& 19.7
\\ \hline $^{13}$N& 5.48$\times $10$^{-2}$(1.00$^{+0.21}_{-0.17}$)& 0.09& 3.4& 2.3
\\ \hline $^{15}$O& 4.80$\times $10$^{-2}$(1.00$^{+0.25}_{-0.19}$)& 0.33& 5.5& 11.8
\\ \hline $^{17}$F& 5.63$\times $10$^{-4}$(1.00$^{+0.25}_{-0.25}$)& 0.0& 0.1& 0.1
\\ \hline Total& & 7.6$^{+1.3}_{-1.1}$& 128$^{+9}_{-7}$& 52.3$^{+6.5}_{-6.0}$
\\ \hline
\end{tabular}
\end{table}
\end{center}
In a next few years even in the optimistic scenario of the
progress with the solar neutrino experiments it will be hardly
possible to decrease this limit lower than 5\% \cite{4} utilizing
only the data on a hydrogen sequence neutrinos because it will be
hardly possible to reach the accuracy in the measurements of these
neutrinos less than 5\%. Lithium experiment has a potential to
confirm or reject the value 1.5\% predicted by the standard solar
model. For a certain energy release in CNO cycle there should be a
corresponding surplus from $^{13}$N- and $^{15}$O-neutrinos to the
total rate expected from the fluxes of neutrinos generated in a
hydrogen sequence alone. The data presented in Table 1 show the
rates calculated by the Standard Solar Model \cite{7} for
different neutrino sources.

One can see from these data that 1$\sigma $ errors vary from 17\%
to 25\% for $^{13}$N- and $^{15}$O-neutrinos close to 1$\sigma $
errors for boron neutrinos. However one should take into account
that there's a correlation of the fluxes of boron and
CNO-neutrinos as it was discussed in \cite{8}. The substantial
issue is that while the contribution of CNO-cycle to the solar
energy is only 1.5\%, the weight of neutrinos from CNO-cycle in
the production rate of $^{7}$Be on $^{7}$Li is about 30\%, see
below. By the time a lithium experiment can start measurements
basically all fluxes but CNO-neutrinos will be measured with
relatively good accuracy and the question about sterile neutrinos
will be cleared to a very small limit, if so. There can be some
delay with the detection of pep-neutrinos, but the ratio of
pep-neutrino flux to pp-neutrino flux is fixed to high accuracy
\cite{1} so it will not present a problem for the evaluation of
the rate from neutrinos of a hydrogen sequence in a lithium
experiment. The effect from solar neutrinos can be measured with
very good accuracy this being a characteristic feature of a
lithium target because the production rate is high and the
cross-section is well known, see Table 1. This is a very rare and
very useful combination for a solar neutrino experiment.

\begin{figure}[!h]
\centering
\includegraphics[width=3in]{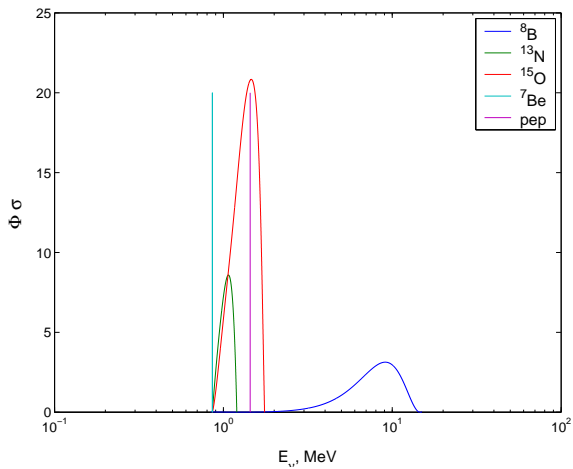}
\caption{The sensitivity plot of lithium detector.}
\end{figure}

\noindent Figure 2 shows the sensitivity plot for a lithium
detector \cite{9}, one can see that the contribution of the
spectra of CNO neutrinos are quite substantial. The present
discovery that MSW LMA region is responsible for the neutrino
oscillations in the Sun means that approximately 1/2 of the
neutrinos with the energy of about 1 MeV and 1/3 of boron
neutrinos coming to the underground detector are of electron type.
Then the total rate expected for a lithium target from solar
neutrinos should be about 23 SNU including the predicted
contribution from CNO-neutrinos about 7 SNU if to take that CNO
cycle produces 1.5\% of the total luminosity of the Sun. Here it
is worth to note that for the present limit 7.3\% the contribution
to the rate of lithium detector from neutrinos of CNO-cycle will
be 35 SNU, more than it is expected from neutrinos of a hydrogen
sequence. This would be soon identified in the running experiment.
Because lithium is a low atomic mass target (very high number of
atoms in 1 g) and the abundance of $^{7}$Li in natural lithium is
93\% even relatively small mass of lithium (10 tons) can provide
high accuracy in the rate measurements. The statistics of a
lithium experiment shows that for the effect of 20 SNU and 4 Runs
per year having the total efficiency of the extraction 80\% and
the efficiency of counting 90\% the statistical accuracy for 4
years of measurements should be about 3-4\%. This is a very good
number quite adequate to determine with good accuracy the
contribution of CNO-neutrinos to the total rate. This will be the
stringent test of the theory of stellar evolution and the
termination of the long-standing goal -- the neutrino spectroscopy
of the interior of the Sun \cite{1}. There's also a potential to
increase the accuracy increasing the total mass of the target
four- eight-fold by using several modules 10 tons each.

It is worth to note that even if the difference of the observed
luminosity of the Sun and the estimated for the hydrogen sequence
is established with good accuracy it does not provide a proof that
the difference is due to CNO-cycle. The proof may be obtained from
another balance - of the measured rates in a lithium detector. The
most interesting thing would be to compare two values: first value
is the contribution of the non-hydrogen sequence to the solar
luminosity found as a difference of the observed luminosity of the
Sun and the one found from the hydrogen sequence alone using the
measured neutrino fluxes of the hydrogen sequence, and second
value is the contribution to the solar luminosity of CNO-cycle
found from the contribution to the measured rate in a lithium
experiment neutrinos of CNO-cycle as a surplus to the rate
determined by a hydrogen sequence. If there would be a substantial
difference of these two values - there is some other source of
energy in the Sun. To make this comparison both values should be
known with good accuracy. As it was shown up the statistics of the
planned lithium experiment enables to get the second value with
good accuracy. What about the first one, the situation here is
probably more complicated. Because the expected contribution of
CNO-cycle to the solar luminosity is only 1.5\% the accuracy in
the neutrino flux measurements from a hydrogen sequence should be
on the level of better than 1\%. Obviously this level will be
reached not soon. So for the nearest future one can talk only
about the contribution of CNO cycle to the total energy production
in the Sun found from a lithium experiment. In fact the
interpretation of the results depends upon how accurately are
measured the parameters of neutrino oscillations. One can expect
that $\Delta $m$^{2}_{12}$ will be found with very good accuracy
by KamLAND in the nearest future. It is not clear yet how fast
will be the progress with the determination of $\Theta _{12}$. If
it will be found with a good accuracy one can investigate the
energy balance

\begin{center}
$L_H$($\Theta _{12}$)+$L_{CNO}$($\Theta _{12}$)=$L_{\odot}$
\end{center}

\noindent to look how accurate is this equality. If on the
contrary the fluxes of the neutrinos of a hydrogen sequence and of
CNO-cycle are measured with very good accuracy while $\Theta
_{12}$ is not one can find $\Theta _{12}$ as the value for which
this equality is fulfilled. Both alternatives look attractive. But
apparently this task is for the future when the flux of
pp-neutrinos will be measured with the accuracy better than 1\%.

A peculiar thing is that although lithium detector is a
radiochemical one i.e. it measures the total rate from all the
neutrino sources on the Sun, there's one possibility to find
separately what is the contribution of $^{13}$N and $^{15}$O
neutrinos. First of all one should note that for lithium detector
the contribution of $^{13}$N neutrinos is 5 times smaller than
that of $^{15}$O neutrinos, see Table 1. It helps in the
interpretation of the results because the interference is small.
But the spectra can be resolved! Let's look more in the details.
If the fluxes of neutrinos from a hydrogen sequence are measured
with very good accuracy then the only unknown thing is the energy
of CNO cycle. But the energy generation in this cycle proceeds by
two half-cycles: from $^{12}$C to $^{14}$N (first one) and from
$^{14}$N to $^{12}$C (second one). The rates depend upon the
abundance of $^{12}$C and $^{14}$N in the interior of the Sun.

\begin{figure}[!h]
\centering
\includegraphics[width=3in]{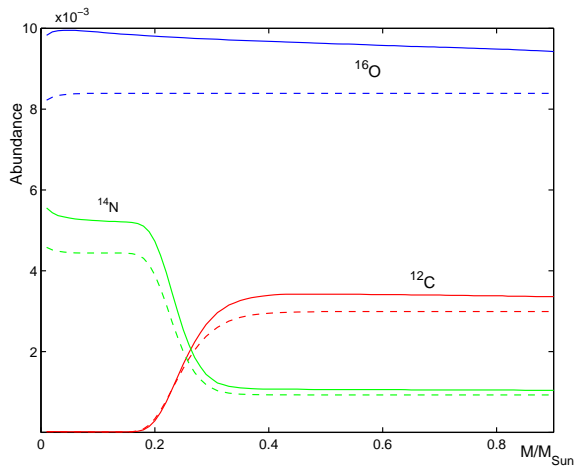}
\caption{The distribution of the abundance of $^{12}C$, $^{14}N$,
$^{16}O$ along the profile of the Sun (in mass ratio units) in SSM
with (solid) and without (dash) helium and metall diffusion
\cite{10}.}
\end{figure}

\noindent Figure 3 shows the distribution of the abundance of
$^{12}$C and $^{14}$N along the profile of the Sun (in a mass
ratio units) \cite{10}. One can see that the center of the Sun is
depleted by $^{12}$C (it is burned out) while is enriched by
$^{14}$N (it is accumulated). The question is: can this abundance
distribution be confirmed by experiment? For the first half-cycle
the energy release E$_{1}$ = $^{12}$C + 2p --$^{14}$N. For the
second one E$_{2}$ = $^{14}$N + 2p -- $^{12}$C -- $\alpha $. The
total energy release will be as it is well know E$_{1}$+ E$_{2}$ =
4p -- $\alpha $. The energy released in the first half-cycle is a
bit smaller than the one in the second half-cycle E$_{2}$--
E$_{1}$ = 2($^{14}$N - $^{12}$C) - $\alpha $. It is about 3.3 MeV.
And if to take into account that the energy of neutrino emitted in
the first half-cycle is less than the energy of the neutrino in
the second one, we obtain that in the first half-cycle the Sun
gets less energy only by about 3.1 MeV than in the second
half-cycle, this means that these energies are very close. What
about the contribution to the rate of lithium detector, the
situation here is very different. The contribution of the
$^{15}$O-neutrino is 5 times bigger than the one of
$^{13}$N-neutrino. Then a system of two equations can be written.

$$ \left\{
\begin{array}{rcl}
L_H+L_{CN}+L_{NO}&=&L_{\odot}\\ R_H+R_{CN}+R_{NO}&=&R_{Li}\\
\end{array}
\right.
 $$

\noindent Here L -- luminosity, R$_{Li}$, R$_{H}$ the measured and
estimated for the hydrogen sequence rates in lithium detector,
R$_{CN}$ and R$_{NO}$ means the rates from neutrino born from
$^{13}$N- and $^{15}$O-decays, R = yL/4$\pi
$R$_{SUN}^{2}\varepsilon $, where R$_{SUN}$ -- the distance from
Sun to Earth, $\varepsilon $ is the energy contributed to the Sun
per one neutrino emitted in each half-cycle of CNO-cycle and y --
the capture rate per one neutrino of $^{13}$N- and
$^{15}$O-spectra. One can see from these equations that
principally it is possible to find separately the fluxes of
$^{13}$N and $^{15}$O neutrinos. The only thing one should know
are the fluxes of neutrinos from a hydrogen sequence and the
parameters of neutrino oscillations. With good accuracy. And of
course to measure the rate by lithium detector.

\begin{figure}[!h]
\centering
\includegraphics[width=3in]{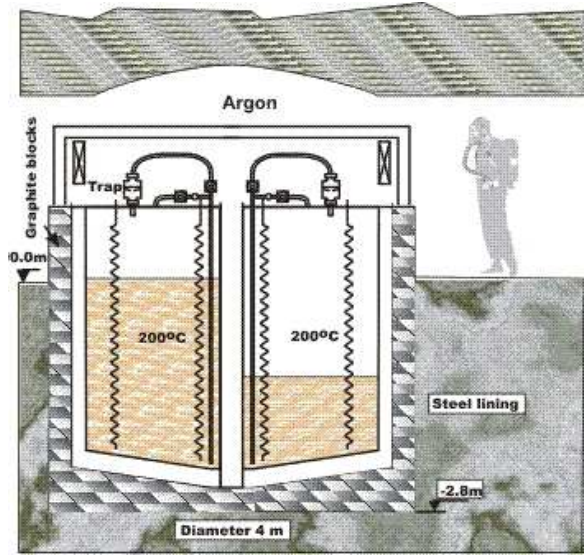}
\caption{The scheme of lithium detector.}
\end{figure}

The technique of lithium experiment is now under development
\cite{10a}. The detector itself can be made quite compact. Figure
4 shows the possible configuration of the detector with one module
of 10 tons of metallic lithium in an underground chamber. One can
see that the scale of the lithium installation is quite modest in
comparison with other solar neutrino detectors. The difficult
point for a lithium project is the counting of $^{7}$Be. To have a
good statistics the efficiency of the counting should be about
80-90\%. But $^{7}$Be decays mainly to the ground state of
$^{7}$Li through the electron capture and the energy of Auger
electron is only 55 eV. It presents a big problem to measure such
a small energy release when one should count single atoms during
long time of measurements (100 days). The decay to the excited
state of $^{7}$Li is accompanied with the gamma-ray of 478 keV
which is a very convenient line for the detection, but the
branching ratio of this mode is only 10\%. The only technique
which looks perspective for a full-scale lithium experiment is a
cryogenic microcalorimetry. The principal possibility of using
this technique for the counting of $^{7}$Be was shown in
\cite{11}, \cite{12}, \cite{13} but for the real technology of
beryllium extraction from lithium and for the low background
environment the appropriate scheme of the detector should be found
and tested.

To summarize we should note that a lithium experiment has a good
discovering potential in the study of solar neutrinos and the more
accurate are the data on the neutrino fluxes from a hydrogen
sequence and on the neutrino oscillation parameters, the more
information on CNO-cycle one can obtain from the results of a
lithium experiment. For a given accuracy of the measurements these
results can be interpreted also in terms of the parameters of
neutrino oscillations, or in terms of the balance violation in the
energies produces in a hydrogen sequence plus CNO-cycle and the
total solar luminosity this being an indication on the other
possible source of solar energy. This work was supported in part
by the Russian Fund of Basic Research, contract N 01-02-16167-A
and by the Leading Russian Scientific School grant N 00-15-96632.

\end{document}